\documentclass{webofc}
\usepackage[varg]{txfonts}   
\usepackage{tikz-feynman}
\tikzfeynmanset{compat=1.1.0}
\usepackage{tikz}
\usepackage{subcaption}

\begin{document}
\title{Thermal modification of open heavy-flavor mesons from an effective hadronic theory }
%
%

\author{\firstname{Gl\`oria} \lastname{Monta\~na}\inst{1}\fnsep\thanks{\email{gmontana@fqa.ub.edu}}
} 

\institute{Departament de F\'isica Qu\`antica i Astrof\'isica and Institut de Ci\`encies del Cosmos (ICCUB), Facultat de F\'isica,  Universitat de Barcelona, Mart\'i i Franqu\`es 1, 08028 Barcelona, Spain 
}

\abstract{%
  We have developed a self-consistent theoretical approach to study the modification of the properties of heavy mesons in hot mesonic matter which takes into account chiral and heavy-quark spin-flavor symmetries. The heavy-light meson-meson unitarized scattering amplitudes in coupled channels incorporate thermal corrections by using the imaginary-time formalism, as well as the dressing of the heavy mesons with the self-energies. We report our results for the ground-state thermal spectral functions and the implications for the excited mesonic states generated dynamically in the heavy-light molecular model. We have applied these to the calculation of meson Euclidean correlators and transport coefficients for $D$ mesons and summarize here our findings. 
}
\maketitle
\section{Introduction}
\label{intro}
Heavy-ion collision experiments in the Relativistic Heavy Ion Collider (RHIC), at BNL, and in the Large Hadron Collider (LHC), at CERN, recreate in the laboratory the conditions for which the quark-gluon plasma (QGP) was created in the early Universe, short after the Big Bang. Fundamental quantities that permit to characterize this medium are thermal spectral and transport properties. 

Heavy quarks, namely the charm and bottom quarks, are produced through the hard-scattering processes at the initial stage of the collisions and interact with the constituents of the QGP.
When the temperature of the system goes below the confining transition temperature heavy quarks hadronize into heavy hadrons, mostly into open heavy-flavor ($D$ and $\bar{B}$) mesons for the high energies achieved at RHIC and LHC, where low baryonic densities are reached. The spectral properties of the open heavy-flavor mesons are modified in the hot hadronic medium from their interaction with the light mesons forming the medium.

Finite temperature corrections to the scattering of $D^{(*)}$ and $D_s^{(*)}$ mesons with the Goldstone bosons ($\pi$, $K$, $\bar{K}$ and $\eta$ mesons) have been implemented within a unitarized thermal approach in \cite{Montana:2020lfi,Montana:2020vjg}. This framework exploits both chiral symmetry and heavy-quark spin-flavor symmetry (HQSFS) to construct the effective Lagrangian that describes the interactions between the heavy and light degrees in vaccuum. HQSFS implies that the interactions are independent of the heavy-quark spin and flavor. The unitarization technique in coupled channels allows us to extend the energy region where the effective theory can be applied, as well as to generate resonances and bound states from the two-body dynamics. 
In this paper we have studied the thermal modification of these states in the hadronic phase, below the critical temperature $T_c$. The finite temperature is incorporated using the imaginary-time formalism (ITF) and the in-medium scattering amplitudes and ground-state self-energies are computed self-consistently.

%

In Section~\ref{sec-1} of this contribution to the proceedings of the Virtual Tribute to Quark Confinement and the Hadron Spectrum 2021 conference we present our latest results on the thermal modification of the properties (masses and widths) of the $D$ and $\bar{B}$ mesons, as well as of the resonances and bound states that are dynamically generated from their scattering with the light mesons. 
The effective theory results for the thermal scattering amplitudes and the spectral functions were used to compute open-charm meson Euclidean correlators in Ref.~\cite{Montana:2020var} and off-shell transport coefficients in Ref.~\cite{Torres-Rincon:2021yga}. The main results are summarized in Sections~\ref{sec-2} and \ref{sec-3} of this contribution, respectively.
Further details can be found in Refs.~\cite{Montana:2020lfi,Montana:2020vjg,Montana:2020var,Torres-Rincon:2021yga}.

\section{Thermal effective field theory for open heavy-flavor mesons}
\label{sec-1}
The interaction in vacuum between the open heavy-flavor mesons ($D^{(*)}=\{D^{(*)},D_s^{(*)}\}$) and the pseudoscalar Goldstone bosons ($\Phi=\{\pi, K, \bar{K}, \eta\}$) is described in Refs.~\cite{Montana:2020lfi,Montana:2020vjg} by means of an effective Lagrangian that is based on chiral symmetry and HQSFS \cite{Guo:2009ct,Liu:2012zya,Tolos:2013kva,Guo:2018tjx}.
The corresponding tree-level amplitudes of the $D\Phi$ (or $D^*\Phi$) scattering for an incoming channel $i$ and an outgoing channel $j$ read
\begin{align} \nonumber\label{eq:potential}
 V^{ij}(s,t,u)=&\ \frac{1}{f_\pi^2}\Big[\frac{C_{\rm LO}^{ij}}{4}(s-u)-4C_0^{ij}h_0+2C_1^{ij}h_1\\ 
 &\ -2C_{24}^{ij}\Big(2h_2(p_2\cdot p_4)+h_4\big((p_1\cdot p_2)(p_3\cdot p_4)+(p_1\cdot p_4)(p_2\cdot p_3)\big)\Big)\\ \nonumber
 &\ +2C_{35}^{ij}\Big(h_3(p_2\cdot p_4)+h_5\big((p_1\cdot p_2)(p_3\cdot p_4)+(p_1\cdot p_4)(p_2\cdot p_3)\big)\Big)
 \Big],
\end{align}
where $s=(p_1+p_2)^2$, $t=(p_1-p_3)^2$ and $u=(p_1-p_4)^2$ are the usual Mandelstam variables. The values of the low-energy constants (LECs) of the terms at NLO in the chiral expansion, $h_{0,...,5}$, were fitted in \cite{Guo:2018tjx} to lattice results of scattering lengths and finite-volum spectra. Keeping lowest order in the heavy-quark mass expansion, there are no tree-level diagrams connecting $D\Phi$ and $D^*\Phi$ channels, and the LECs take the same values for these two sectors. The isospin coefficients $C_{ij}$ for the different strangeness and isospin sectors and the values of the LECs are given in Tables~III and IV in Ref.~\cite{Montana:2020vjg}. Relying on heavy-quark flavor symmetry (HQFS) the interaction potential of Eq.~(\ref{eq:potential}) can be extended to the open-bottom sector upon the replacement $D\rightarrow\bar{B}$ and the rescaling of the LECs with the heavy meson masses, 
\begin{equation} 
 \frac{h_{i}^D}{\hat{m}_D}=\frac{h_{i}^B}{\hat{m}_B} \ , \quad \textrm{for} \; i=\{0,1,2,3\} \ , \quad \textrm{and} \quad
 h_{4,5}^D\hat{m}_D=h_{4,5}^B\hat{m}_B \ ,  \quad \textrm{for} \; i=\{4,5\} \ .
\end{equation}

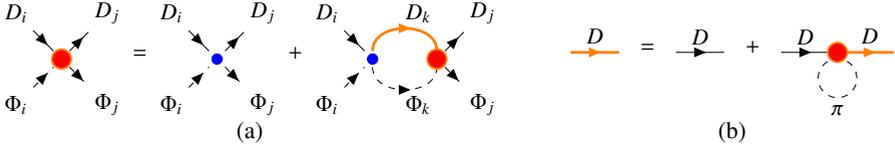
\begin{figure}[htbp!]
\centering 
\begin{subfigure}[b]{0.54\textwidth} \centering
\captionsetup{skip=-2pt}
 \scalebox{0.85}{\begin{tikzpicture}[baseline=(i.base)]
    \begin{feynman}[small]
      \vertex (a) {\(D_i\)};
      \vertex [below right = of a] (i) {};
      \vertex [above right = of i] (b) {\(D_j\)};
      \vertex [below right = of i] (d) {\(\Phi_j\)};
      \vertex [below left=of i] (c) {\(\Phi_i\)};
      \diagram* {
        (a) -- [fermion] (i), 
        (i) -- [fermion] (b),
        (c) -- [charged scalar] (i),
        (i) -- [charged scalar] (d),
       };
     \draw[dot,minimum size=4mm,thick,orange,fill=red] (i) circle(1.5mm);
    \end{feynman}
  \end{tikzpicture}
  $=$
  \begin{tikzpicture}[baseline=(i.base)]
    \begin{feynman}[small]
      \vertex (a) {\(D_i\)};
      \vertex [below right = of a] (i) {};
      \vertex [above right = of i] (b) {\(D_j\)};
      \vertex [below right = of i] (d) {\(\Phi_j\)};
      \vertex [below left=of i] (c) {\(\Phi_i\)};
      \diagram*{
        (a) -- [fermion] (i), 
        (i) -- [fermion] (b),
        (c) -- [charged scalar] (i),
        (i) -- [charged scalar] (d),
       } ;    
     \draw[dot,blue,fill=blue] (i) circle(.8mm);
    \end{feynman}
  \end{tikzpicture}
  $+$
  \begin{tikzpicture}[baseline=(i.base)]
    \begin{feynman}[small]
      \vertex (a) {\(D_i\)};
      \vertex [below right = of a] (i) {};
      \vertex [right = of i] (j) {};
      \vertex [above right = of j] (b) {\(D_j\)};
      \vertex [below right = of j] (d) {\(\Phi_j\)};
      \vertex [below left=of i] (c) {\(\Phi_i\)};
      \vertex [above right = of i] (b1) {\(D_k\)};
      \vertex [below right=of i] (d1) {\(\Phi_k\)};
      \diagram*{
        (a) -- [fermion] (i), 
        (j) -- [fermion] (b),
        (i) -- [orange, fermion, very thick, half left, looseness=1.2] (j),
        (i) -- [charged scalar, half right, looseness=1.2] (j),
        (c) -- [charged scalar] (i),
        (j) -- [charged scalar] (d),
       } ;    
     \draw[dot,blue,fill=blue] (i) circle(0.8mm);
     \draw[dot,minimum size=4mm,thick,orange,fill=red] (j) circle(1.5mm);
    \end{feynman}
  \end{tikzpicture}}
\caption{}
\label{fig:BS-a}
\end{subfigure}
\hspace{3mm}\begin{subfigure}[b]{0.38\textwidth}\centering
\captionsetup{skip=-2pt}
 \scalebox{0.85}{\begin{tikzpicture}[baseline=(a.base)]
    \begin{feynman}[small]
      \vertex (a) {};
      \vertex [right = of a] (b) {};
      \diagram* {
        (a) -- [orange, fermion, very thick,edge label=\(\textcolor{black}{D}\)] (b), 
       };
    \end{feynman}
  \end{tikzpicture}
  $=$
 \begin{tikzpicture}[baseline=(a.base)]
    \begin{feynman}[small]
      \vertex (a) {};
      \vertex [right = of a] (b) {};
      \diagram* {
        (a) -- [fermion, edge label=\(D\)] (b), 
       };
    \end{feynman}
  \end{tikzpicture}
  $+$
  \begin{tikzpicture}[baseline=(a)]
    \begin{feynman}[small, inline=(a)]
      \vertex (a) {};
      \vertex [right = of a] (i) {};
      \vertex [right = of i] (b) {};
      \vertex [below = 0.4cm of i] (d) {};
      \vertex [below = 0.9cm of i] (e) {\(\pi\)};
      \diagram* {
        (a) -- [fermion,edge label=\(D\)] (i), 
        (i) -- [orange, fermion, very thick,edge label=\(\textcolor{black}{D}\)] (b),
       } ;   
     \draw[dashed] (d) circle(0.3cm);
     \draw[dot,minimum size=4mm,thick,orange,fill=red] (i) circle(1.5mm);
    \end{feynman}
  \end{tikzpicture}}
\caption{}
\label{fig:BS-b}
\end{subfigure} \vspace{-0.5cm}
\caption{(a) Bethe-Salpeter equation and (b) Dyson-equation for the heavy-meson propagator at finite temperature. Big red circles represent the $T$-matrix, small blue circles correspond to the interaction kernel and thick orange lines depict dressed propagators. Figure taken from Ref.~\cite{Montana:2020lfi}.}\vspace{-0.5cm}
\label{fig:BS}
\end{figure}

The $s$-wave projection of the amplitude in Eq.~(\ref{eq:potential}) is unitarized by solving the Bethe-Salpeter equation in coupled channels (Fig.~\ref{fig:BS-a}),
\begin{equation}\label{eq:BS}
 T_{ij}(s)=V_{ij}(s)+V_{ik}(s)G_kT_{kj}(s) \ ,
\end{equation}
where $T$ is the unitarized amplitude, $V$ is the $s$-wave interaction kernel and $G$ is the loop function constructed from the two meson propagtors,
\begin{equation}
 \label{eq:loopVac}
  G^l(s)=i\int\frac{d^4q}{(2\pi)^4} \frac{1}{q^2-m_D^2+i\epsilon} \frac{1}{(p-q)^2-m_\Phi^2+i\epsilon} \ ,
\end{equation}
which is regularized with a hard momentum cutoff $\Lambda$. Equation~(\ref{eq:BS}) has an algebraic solution in the on-shell approximation,
$ T(s)=V(s)[1-V(s)G(s)]^{-1}$,
which is a matrix equation for the case of two or more coupled channels.

At finite temperature the loop function $G$ is calculated through the ITF, in which the time dimension is Wick rotated and compactified in the range $[0,\beta=1/T]$, with $T$ the temperature of the system, and the energy integral transforms into a summation over discrete Matsubara frequencies, $q^0\rightarrow i2\pi nT$ for bosons \cite{Das:1997gg,Weldon:1983jn}. We also introduce the Lehmann representation for the meson propagators
\begin{equation}
 \mathcal{D}_M(i\omega_n,\textbf{q})=\int d\omega'\frac{S_M(\omega',\textbf{q})}{i\omega_n-\omega'} \ ,
\end{equation}
where $S_M$ is the spectral function and $M$ denotes the meson species ($D$ or $\Phi$). After the Matsubara summation and the analytical continuation to real energies, $i\omega_m\rightarrow E+i\varepsilon$, the thermal loop function reads
\begin{equation}\label{eq:loop}
    G_{D\Phi}(E,\textbf{p};T)=\int\frac{d^3q}{(2\pi)^3}\int d\omega\int d\omega'\frac{S_{D}(\omega,\textbf{q};T)S_{\Phi}(\omega',\textbf{p}-\textbf{q};T)}{E-\omega-\omega'+i\varepsilon}[1+f(\omega,T)+f({\omega'},T)] \ ,
\end{equation}
where $f(\omega,T)=[\exp(\omega/T)-1]^{-1}$ is the equilibrium Bose-Einstein distribution function. 

The spectral function of the heavy meson is given by
\begin{equation}\label{eq:S}
    S_{D}(\omega,\textbf{q};T)=-\frac{1}{\pi}{\rm Im\,}\mathcal{D}_{D}(\omega,\textbf{q};T)=-\frac{1}{\pi}{\rm Im\,}\Bigg(\frac{1}{\omega^2-\textbf{q}^2-m_{D}^2-\Pi_{D}(\omega,\textbf{q};T)}\Bigg) \ ,
\end{equation}
where the self-energy follows from closing the light-meson line in the corresponding $T$-matrix element (Fig.~\ref{fig:BS-b}),
\begin{equation}\label{eq:pi}
    \Pi_D(E,\textbf{p};T)=\frac{1}{\pi}\int\frac{d^3q}{(2\pi)^3}\int d\Omega\frac{E}{\omega_\Phi}\frac{f(\Omega,T)-f(\omega_\Phi,T)}{E\,^2-(\omega_\Phi-\Omega)^2+i\varepsilon}{\rm Im\,}T_{D\Phi}(\Omega,\textbf{p}+\textbf{q};T) \ .
\end{equation}
In Refs.~\cite{Montana:2020lfi,Montana:2020vjg} we have approximated the spectral function of the light meson to a $\delta$-type spectral function because the in-medium modification of the pions is small. At $T\lesssim 150\rm\,MeV$ the largest contribution to the self-energy comes from the interaction of the heavy mesons with the pions, as the abundance of heavier light mesons, i.e. kaons, antikaons and eta mesons, is thermally suppressed. This has been discussed in detail in Ref.~\cite{Montana:2020vjg}. In the results below only the pion contribution is considered in Eq.~(\ref{eq:pi}) to calculate the heavy-meson self-energy.

\begin{figure}[htbp!]
\centering
\includegraphics[width=\textwidth]{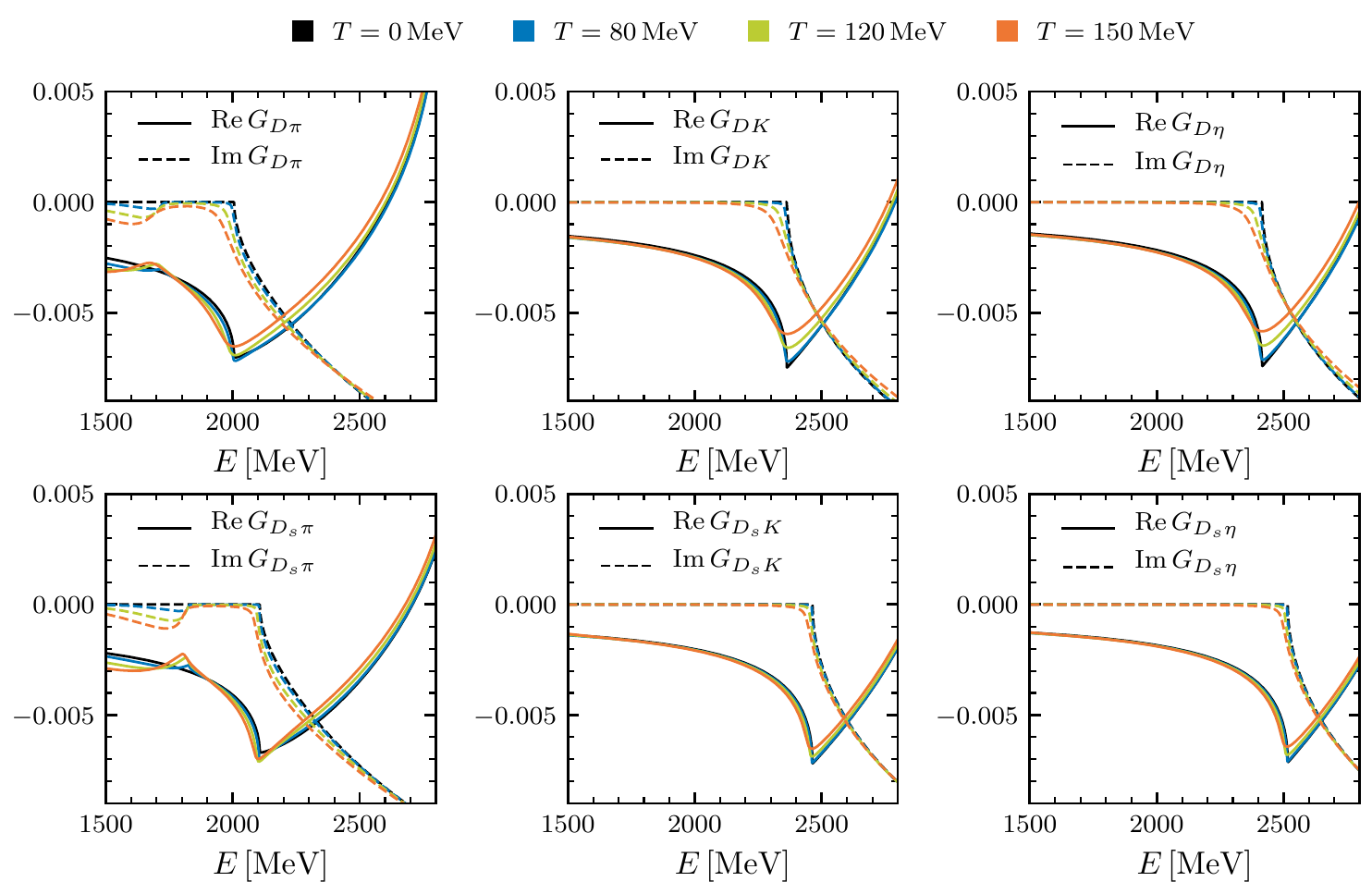}\vspace{-0.2cm}
\caption{Real and imaginary parts of the two-meson loop function at finite temperature. Top plots: $D\Phi$ case. Bottom plots: $D_s\Phi$ case. Data taken from \cite{Montana:2020vjg}.}\vspace{-0.5cm}
\label{fig:loops_ps}       
\end{figure}

Equations~(\ref{eq:loop}) to (\ref{eq:pi}) are coupled to each other and require a self-consistent solution. The results obtained for the loop function for $D\Phi$ and $D_s\Phi$ channels are shown in the top and bottom panels of Fig.~\ref{fig:loops_ps}, respectively, for different temperatures. At $T=0$ (black lines) the imaginary part becomes non-zero in the energy region above threshold ($m_{D/D_s}+m_\Phi$), where the unitary cut develops. At $T>0$ an additional cut opens up for energies below $m_{D/D_s}-m_\Phi$. This is the so-called Landau cut and it is related to thermal production and absorption processes \cite{Das:1997gg,Weldon:1983jn}. From the loop functions in Fig.~\ref{fig:loops_ps} it is clear that the Landau cut is much larger for channels involving pions (for $K$ and $\eta$ mesons it cannot be distinguished with the scale used for the plots), and its importance grows with increasing temperatures because of the larger abundance of thermal pions. In addition, at finite temperature, the boundaries of the cuts are smoothened due to the widening of the heavy-meson spectral function, as discussed in \cite{Montana:2020lfi,Montana:2020vjg}. 

\begin{figure}[hbtp!]
\centering
\includegraphics[width=\textwidth,clip]{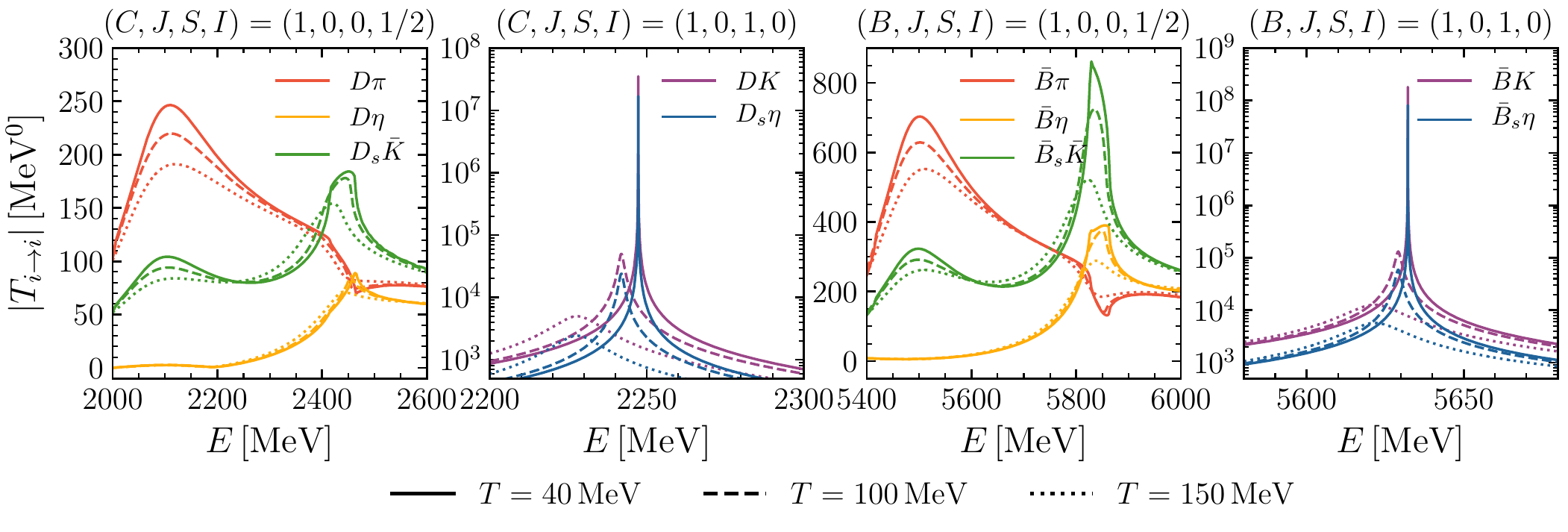}\vspace{-0.4cm}
\caption{$T$-matrix at finite temperature for the diagonal transitions in the sectors with charm $C=1$ (bottom $B=1$), spin $J=0$ and strangeness and isospin $(S,I)=\{(0,1/2),(1,0)\}$. Figure extracted from \cite{Montana:2021vks} with data partially taken from \cite{Montana:2020lfi}.}\vspace{-0.2cm}
\label{fig:Tmatrix}       
\end{figure}

Figure~\ref{fig:Tmatrix} shows the self-consistent results for the unitarized scattering amplitudes at finite temperature in the sectors in which dynamically generated states with $J^P=0^+$ appear. In the charm sector with strangeness and isospin $(S,I)=(0,1/2)$ (first panel) the two resonances are identified with the experimental $D_0^*(2300)$ state, for which a two-pole structure at $T=0$ was described in vaccuum in \cite{Albaladejo:2016lbb}. The lower resonance is seen in the $D\pi$ scattering amplitude, while the higher one is visible in the $D_s\bar{K}$ channel. The thermal effects on the generation of these states are moderate and the structures are still clearly seen at $T=150~\textrm{MeV}$. In the sector with $(S,I)=(1,0)$ (second panel) the bound state associated to the $D_{s0}^*(2317)$ in vacuum gains a substantial thermal width as the temperature is increased. The bottom counterparts of these states have not been reported experimentally yet. We predict here their thermal modification to be of the same order as in the charm sector, as expected from HQFS (third and fourth panels). The thermal scattering amplitudes for $D^*\Phi$ and $D_s^*\Phi$ and the corresponding resonances with $J^P=1^+$ are reported in \cite{Montana:2020vjg}.

\begin{figure}[htbp!]
\centering
\includegraphics[width=11cm,clip]{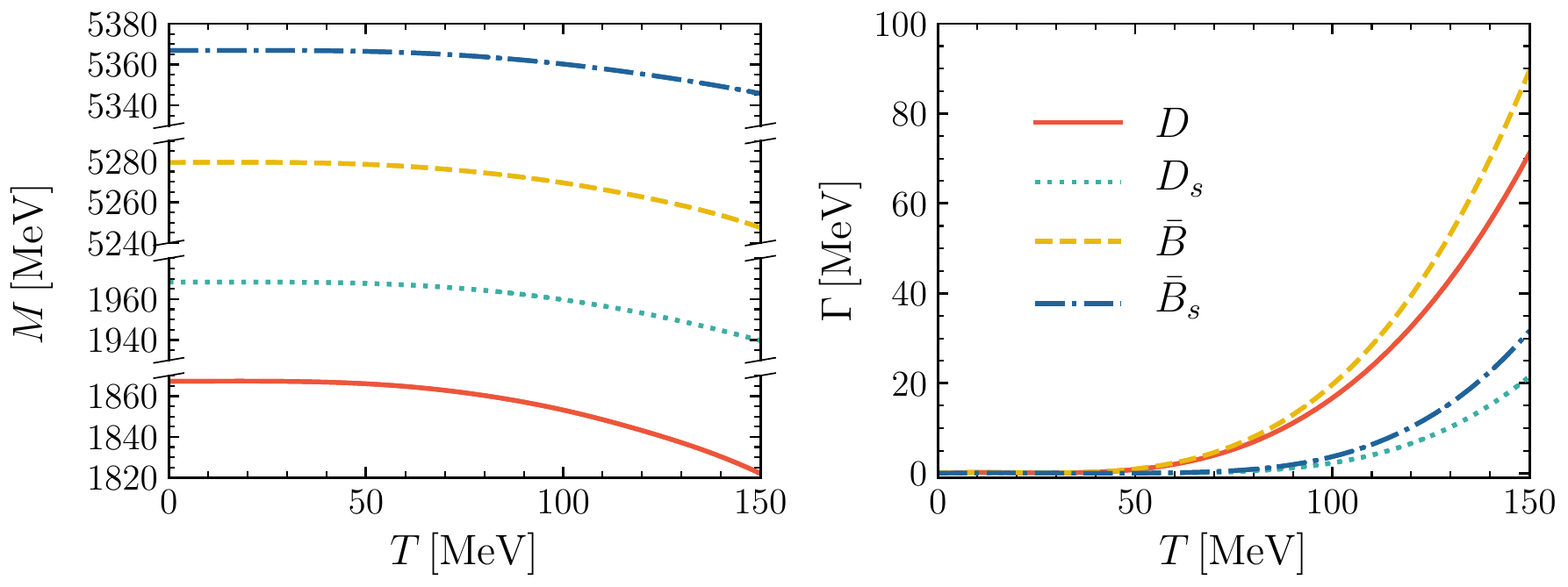}\vspace{-0.2cm}
\caption{Thermal evolution of the mass (left panel) and the width (right panel) of the open heavy-flavor ground states.  Data partially taken from \cite{Montana:2020vjg}.}
\label{fig:M_Width_gs}       
\end{figure}

The thermal modification of the ground states $D$, $D_s$, $\bar{B}$ and $\bar{B}_s$ is extracted from their respective spectral functions. Figure~\ref{fig:M_Width_gs} displays the evolution of the masses and the widths as a function of temperature in the left and right panels, respectively.
The heavy ground states, which are stable degrees of freedom of the model in vacuum, acquire a thermal width due to their interaction with the thermal pions of the bath. Furthermore the masses decrease of the order of $\sim 20-50$ MeV at the highest temperature considered ($T=150~\textrm{MeV}$).
The thermal modification of the open-charm vector counterparts is given in \cite{Montana:2020vjg}. 

\section{Open-charm Euclidean correlators}
\label{sec-2}
In this section we present the results for the Euclidean meson correlators of open-charm mesons that were obtained in our previous work~\cite{Montana:2020var}.
Temporal Euclidean meson correlators in a hot medium can be computed on the lattice for a discrete number of points in Euclidean time $\tau$. The temperature of the system is related to the temporal extent of the lattice, $T=1/(aN_\tau)$.
The spectral function is related to the Euclidean correlator through the integral relation:
\begin{equation}\label{eq:corr}
    G_E(\tau;T)=\int_0^\infty d\omega\, K(\tau,\omega;T)\,\rho(\omega;T) \ , 
\end{equation}
where the spectral function $\rho(\omega;T)$ is convoluted with a temperature dependent kernel,
\begin{equation}
 K(\tau,\omega;T)=\cosh\Bigg[\omega\Bigg(\tau-\frac{1}{2T}\Bigg)\Bigg]\sinh^{-1}\Bigg(\frac{\omega}{2T}\Bigg) \ .
\end{equation}

The extraction of a continuous spectral function from the lattice QCD correlators requires the inversion of Eq.~(\ref{eq:corr}). However such an inversion of a discrete number of data points with statistical errors is an ill-posed problem and the methods that are usually employed rely on specific assumptions for the spectral function. In Ref.~\cite{Montana:2020var} we took the opposite approach. In order to compare the effective field theory results with the lattice QCD calculations for open-charm mesons at finite temperature of Ref.~\cite{Kelly:2018hsi} we insert the thermal spectral function in Eq.~(\ref{eq:corr}) and obtain the corresponding Euclidean correlator.

 The full meson spectral function $\rho(\omega;T)$ contains information of the ground state and the additional contribution of excited bound states and the continuum of scattering states. The ground-state spectral function is obtained using Eq.~(\ref{eq:S}) and the continuum is taken into account through a parametrization of the free meson spectral function in the non-interacting limit, while the contribution of the possible excited states is neglected in \cite{Montana:2020var}. Thus, the full spectral function reads
 \begin{equation}\label{eq:fullS}
    \rho(\omega;T)= M_D^4S_D(\omega;T)+a\rho_{\rm cont}(\omega;T) \ ,
\end{equation}
where the contribution of the continuum is weighted with respect to the ground-state contribution through the factor $a$.

\begin{figure}[htbp!]
\centering
\includegraphics[width=11cm,clip]{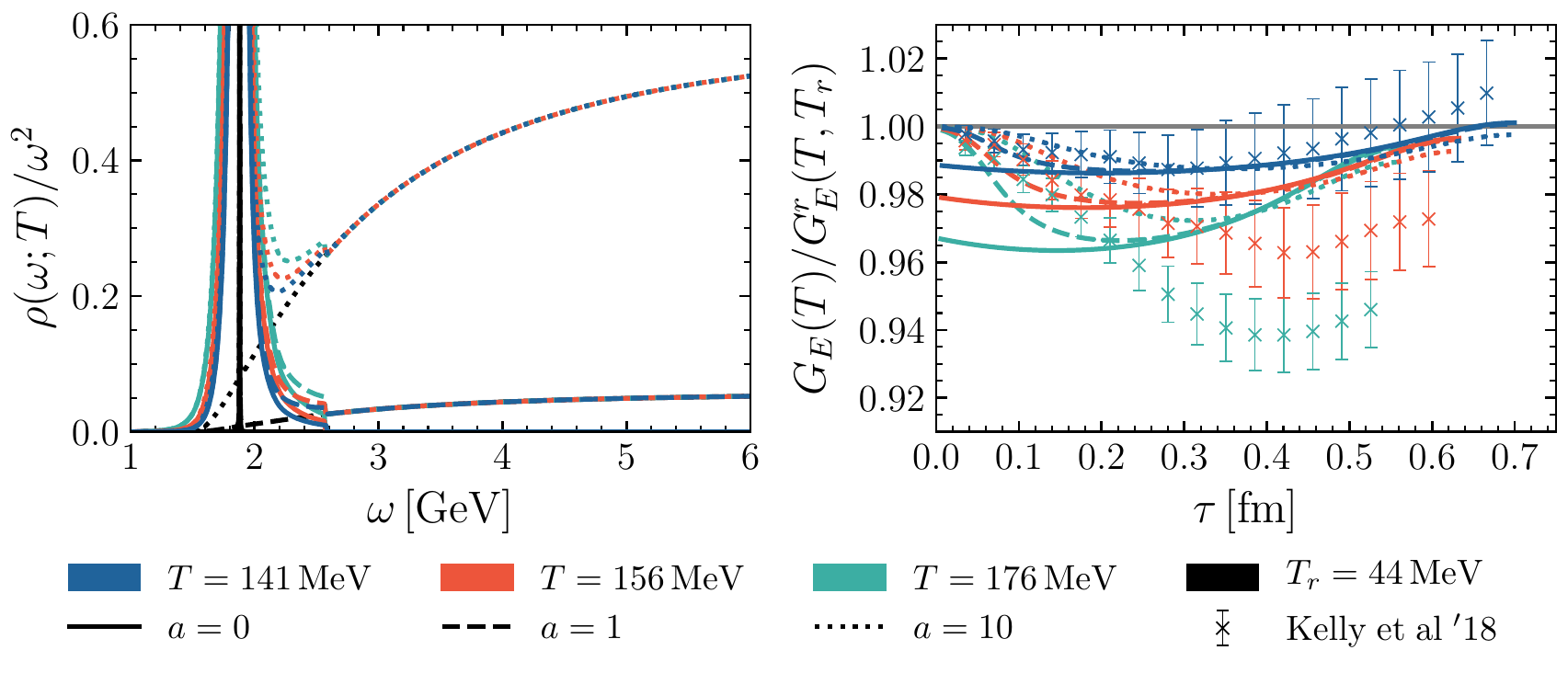}\vspace{-0.2cm}
\caption{Left panel: Full spectral functions of the $D$-meson for unphysical meson masses. Right panel: Ratio of the Euclidean correlator and the reconstructed correlator at $T_r=44~\textrm{MeV}$. The results are shown for temperatures below $T_c=185~\textrm{MeV}$ \cite{Kelly:2018hsi} and for various values of the weight parameter $a$. Figure extracted from \cite{Montana:2021vks} with data taken from \cite{Montana:2020var}.}
\label{fig:correlators}       
\end{figure}

The full spectral functions obtained for the $D$ mesons in a thermal medium of  pions of mass $m_\pi=384~\textrm{MeV}$, i.e. the unphysical pion mass used in the lattice calculations of Ref.~\cite{Kelly:2018hsi}, are shown in the left panel of Fig.~\ref{fig:correlators}. The results are given for various temperatures below the pseudocritical temperature on the lattice ($T_c=185~\textrm{MeV}$) and for the weight parameter $a=\{0,1,10\}$ in Eq.~(\ref{eq:fullS}).

In order to discern the modification of the correlator due to the temperature dependence of the spectral function alone and cancel that of the integration kernel, the ratio with a recostructed correlator is usually computed. This is shown in the right panel of Fig.~\ref{fig:correlators}, together with the results of the lattice simulations of Ref.~\cite{Kelly:2018hsi}. The inclusion of the continuum improves the behavior of the ratio at small $\tau$ and the results are in good agreement with the lattice QCD data within the error bars for the lowest temperatures considered.

\section{$D$-meson transport coefficients}
\label{sec-3}
In this section we summarize our latest results on open-charm transport coefficients of Ref.~\cite{Torres-Rincon:2021yga}, where we implemented the thermal scattering amplitudes and the $D$-meson spectral functions in order to take into account thermal and off-shell effects. After some justified approximations, the off-shell kinetic equation derived in \cite{Torres-Rincon:2021yga} following the Kadanoff-Baym approach~\cite{kadanoff1962quantum} reduces to an off-shell Fokker-Planck equation for the Green's function,
\begin{equation}\label{eq:FokkerPlanck}
    \frac{\partial}{\partial t} G_D^< (t,k) = \frac{\partial}{\partial k^i} \left\{ \hat{A} (k;T) k^i G_D^< (t,k) + \frac{\partial}{\partial k^j} \left[ \hat{B}_0(k;T) \Delta^{ij} + \hat{B}_1(k;T) \frac{k^i k^j}{|{\textbf{k}|}^2} \right] G_D^< (t,k) \right\} \ .
\end{equation}
In this equation the drag force $\hat{A}$ and the diffusion coefficients $\hat{B}_0$ and $\hat{B}_1$ are defined off shell, as they are calculated using spectral functions, as well as thermal scattering amplitudes. They are defined in \cite{Torres-Rincon:2021yga} as
\begin{align}
 \hat{A}(k^0,\textbf{k};T)&\equiv \bigg\langle 1-\frac{\textbf{k}\cdot\textbf{k}_1}{\textbf{k}^2}\bigg\rangle \ , \\
 \hat{B}_0(k^0,\textbf{k};T)&\equiv \frac14\bigg\langle \textbf{k}_1^2-\frac{(\textbf{k}\cdot\textbf{k}_1)^2}{\textbf{k}^2}\bigg\rangle \ , \\ 
 \hat{B}_1(k^0,\textbf{k};T)&\equiv \frac12\bigg\langle \frac{[\textbf{k}\cdot(\textbf{k}-\textbf{k}_1)]^2}{\textbf{k}^2}\bigg\rangle \ ,
\end{align}
where $\langle \mathcal{F}\rangle$ stands for the average of the quantity $\mathcal{F}$ weighted with the heavy-meson spectral function, the $T$-matrix and the appropriate Bose-Einstein factors. Full details of the calculation and the results for these transport coefficients are given in \cite{Torres-Rincon:2021yga}.

\begin{figure}[htbp!]
\centering
\sidecaption
\includegraphics[width=7cm,clip]{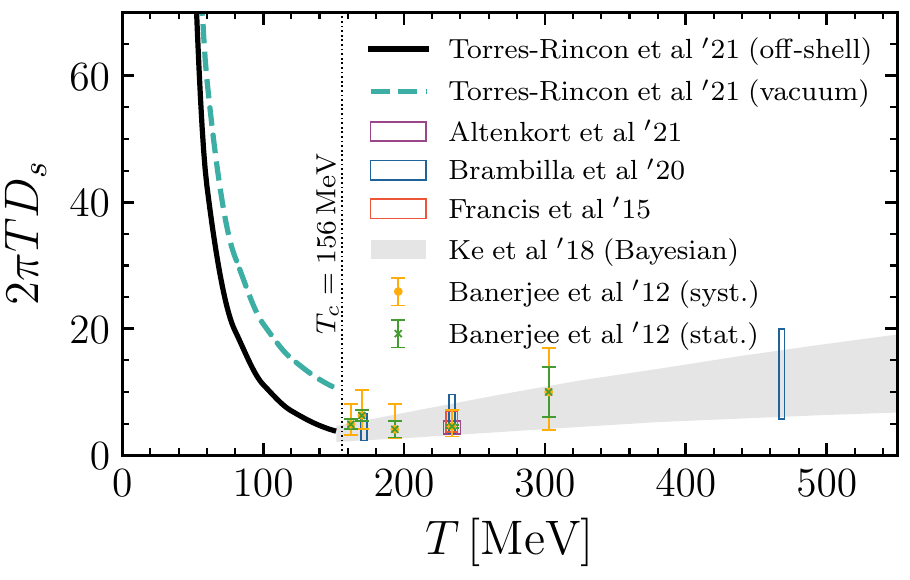}
\caption{Spatial diffusion coefficient below and above $T_c$ resulting from different approaches (see \cite{Torres-Rincon:2021yga} for details). The empty boxes correspond to a range of values of $2\pi TD_s$ for the central value of $T$. Figure extracted from \cite{Montana:2021vks} with data taken from \cite{Torres-Rincon:2021yga}. }
\label{fig:transport}       
\end{figure}

The spatial diffusion coefficient $D_s$ is related to $\hat{B}_0$ in the static limit,
\begin{equation}
    2\pi TD_s(T)=\lim_{{\textbf{k}}\rightarrow 0}\frac{2\pi T^{\,3}}{\hat{B}_0({E_k,\textbf{k}};T)} \ ,
\end{equation}
where $E_k$ is the quasiparticle energy. The result for this transport coefficient is shown in Fig.~\ref{fig:transport} below $T_c$ for the case of the off-shell thermal calculation (black solid line) and it is compared with the on-shell approximation (green dashed line), in which the spectral function is approximated to a $\delta$ and the thermal $T$-matrix is replaced by the corresponding one in vacuum. The main difference is due to the contribution of the Landau cut when thermal scattering amplitudes are employed. We stress that, when thermal corrections are included, the $D$-meson spatial diffusion coefficient matches smoothly the recent lattice QCD calculations and Bayesian analyses performed above $T_c$.

\section{Conclusions}
In this contribution we have presented our recent results for heavy mesons at finite temperature, including the calculation of Euclidean meson correlators and transport coefficients. We have described the interactions of heavy mesons in a bath of light mesons using a thermal effective field theory based on chiral and heavy quark symmetries. The thermal corrections to the heavy-meson self-energies and the unitarized amplitudes have been obtained in a self-consistent manner. We have shown recent results of the impact that the thermal medium has on the properties of the ground-states (masses and widths) as well as the on the states generated dynamically from the two-meson interaction.

Our results for the Euclidean correlators are compatible with the existing lattice QCD data below $T_c$. It would be interesting to compare both approaches at smaller temperatures and meson masses closer to the physical values, as well as to apply techniques for the reduction of excited-state contamination from the correlators in the analysis of the lattice QCD data. From the effective theory side, one might need to include the contribution of the thermal kaons to the $D$-meson self-energy, specially taking into account the reduction of the mass gap between kaons and pions on the lattice. 

Finally, we have shown that the use of thermal scattering amplitudes to compute charm transport coefficients entails the appearance of a new term related to the Landau cut of the $T$-matrix at finite temperature. This contribution is absent if vacuum scattering amplitudes are employed, as commonly done in the past. In particular we have seen that its consideration gives rise to a smooth matching of the transport coefficents with lattice QCD calculations and Bayesian analyses around $T_c$.

\begin{acknowledgement}

This work has been supported by the projects CEX2019-000918-M (Unidad de Excelencia ``Mar\'{\i}a de Maeztu"), PID2019-110165GB-I00 and PID2020-118758GB-I00 financed by MCIN/AEI/10.13039/501100011033. The author also acknowledges support from the FPU17/04910 Doctoral Grant from the Spanish Ministerio de Ciencia Innovación y Universidades.
\end{acknowledgement}

%
\bibliography{proc_vConf21_GM}

\begin{thebibliography}{14}

\bibitem{Montana:2020lfi}
G.~Monta\~na, A.~Ramos, L.~Tolos, J.M. Torres-Rincon, Phys. Lett. B
  \textbf{806}, 135464 (2020), \texttt{2001.11877}

\bibitem{Montana:2020vjg}
G.~Monta\~na, A.~Ramos, L.~Tolos, J.M. Torres-Rincon, Phys. Rev. D
  \textbf{102}, 096020 (2020), \texttt{2007.12601}

\bibitem{Montana:2020var}
G.~Monta\~na, O.~Kaczmarek, L.~Tolos, A.~Ramos, Eur. Phys. J. A \textbf{56},
  294 (2020), \texttt{2007.15690}

\bibitem{Torres-Rincon:2021yga}
J.M. Torres-Rincon, G.~Monta\~na, A.~Ramos, L.~Tolos (2021),
  \texttt{2106.01156}

\bibitem{Guo:2009ct}
F.K. Guo, C.~Hanhart, U.G. Meissner, Eur. Phys. J. \textbf{A40}, 171 (2009),
  \texttt{0901.1597}

\bibitem{Liu:2012zya}
L.~Liu, K.~Orginos, F.K. Guo, C.~Hanhart, U.G. Meissner, Phys. Rev.
  \textbf{D87}, 014508 (2013), \texttt{1208.4535}

\bibitem{Tolos:2013kva}
L.~Tolos, J.M. Torres-Rincon, Phys. Rev. \textbf{D88}, 074019 (2013),
  \texttt{1306.5426}

\bibitem{Guo:2018tjx}
Z.H. Guo, L.~Liu, U.G. Mei{\ss}ner, J.A. Oller, A.~Rusetsky, Eur. Phys. J.
  \textbf{C79}, 13 (2019), \texttt{1811.05585}

\bibitem{Das:1997gg}
A.K. Das, \emph{{Finite Temperature Field Theory}} (World Scientific, New York,
  1997), ISBN 9789810228569, 9789814498234

\bibitem{Weldon:1983jn}
H.A. Weldon, Phys. Rev. \textbf{D28}, 2007 (1983)

\bibitem{Albaladejo:2016lbb}
M.~Albaladejo, P.~Fernandez-Soler, F.K. Guo, J.~Nieves, Phys. Lett.
  \textbf{B767}, 465 (2017), \texttt{1610.06727}

\bibitem{Montana:2021vks}
G.~Montana, A.~Ramos, L.~Tolos, J.M. Torres-Rincon, \emph{{Temperature
  dependence of the properties of open heavy-flavor mesons}}, in \emph{{19th
  International Conference on Strangeness in Quark Matter}} (2021),
  \texttt{2108.04874}

\bibitem{Kelly:2018hsi}
A.~Kelly, A.~Rothkopf, J.I. Skullerud, Phys. Rev. D \textbf{97}, 114509 (2018),
  \texttt{1802.00667}

\bibitem{kadanoff1962quantum}
{Kadanoff, L.P. and Baym, G.}, \emph{{Quantum statistical mechanics}} ({New
  York: WA Benjamin}, {1962})

\end{thebibliography}
%
%

\end{document}